\documentclass[acmsmall]{acmart}

\AtBeginDocument{%
  \providecommand\BibTeX{{%
    \normalfont B\kern-0.5em{\scshape i\kern-0.25em b}\kern-0.8em\TeX}}}

\setcopyright{none}
\copyrightyear{2022}
\acmYear{2022}
\acmDOI{XXXXXXX.XXXXXXX}

\begin{document}

\title{ECHO : An Automated Contextual Inquiry Framework for Anonymous Qualitative Studies using Conversational Assistants}




\author{Rishika Dwaraghanath}
\affiliation{%
  \institution{Shiv Nadar University}
  \city{Greater Noida}
  \state{Uttar Pradesh}
  \country{India}}
\email{rt347@snu.edu.in}

\author{Rahul Majethia}
\affiliation{%
  \institution{Shiv Nadar University}
  \city{Greater Noida}
  \state{Uttar Pradesh}
  \country{India}}
\email{rahul.majethia@snu.edu.in}

\author{Sanjana Gautam}
\affiliation{%
  \institution{Pennsylvania State University}
  \city{State College}
  \state{Pennsylvania}
  \country{USA}}
\email{sqg5699@psu.edu}



\begin{abstract}
Qualitative research studies often employ a contextual inquiry, or a field study that involves in-depth observation and interviews of a small sample of study participants, in-situ, to gain a robust understanding of the reasons and circumstances that led to the participant’s thoughts, actions, and experiences regarding the domain of interest. Contextual inquiry, especially in sensitive data studies, can be a challenging task due to reasons such as participant privacy, as well as physical constraints such as in-person presence and manual analysis of the qualitative data gathered. In this work, we discuss Enquête Contextuelle Habile Ordinateur (ECHO); a virtual-assistant framework to automate the erstwhile manual process of conducting contextual inquiries and analysing the respondents’ subjective qualitative data. ECHO automates the contextual inquiry pipeline, while not compromising on privacy preservation or response integrity. Its adaptive conversational interface enables respondents to provide unstructured or semi-structured responses in free-form natural language, allowing researchers to explore larger narratives in participant response data. ECHO also supports response-driven exploratory questions and automates coding methodologies for qualitative data, thus enabling the inquirer to dive deeper into correlated questions and to do better cause-effect analysis. It focuses on addressing the limitations of manual response annotation, bringing standardisation to free-form text, and eliminating perspective bias amongst different reviewers of subjective responses. A participatory mental health study was conducted on 167 young adults bifurcated into two focus groups; one of which was administered a conventional contextual inquiry, and the other via ECHO, virtually. ECHO outperformed on participant transparency, response detail and trivially, median time required for end-to-end inquiry completion, per participant.

\end{abstract}


\maketitle

\section{Introduction}
\label{sec:introduction}
Ethnographic research methodologies have been a fundamental tool in observing and studying human behaviour in diverse fields — healthcare, technology, anthropology etc. \cite{intro_1,intro_3}. Broadly, ethnography can be described as the study of a group of people with a common factor for which they are being observed or interviewed for in their natural environment, by engaging in activities with the group. Early examples of social anthropology studies point to prominent researchers like Margaret Mead who believed that only by living and experiencing daily life with the tribes could they gain a real understanding of the natives’ culture and way of life \cite{intro_1, intro_2}. Thus, the goal of ethnography is to perceive activities as social acts that take place within a socially organised domain and are carried out through and via the individuals' daily activities. The researcher’s presence at the study participant's natural surroundings, as well as the extensive observation of events, practises, dialogues, and activities that make up its foundation for being from the participants’ point of view, are said to be its defining characteristics.

Contextual inquiries \cite{intro_6} comprise an important subset of ethnographic research which share the same core ideology. But in addition to observing and understanding the participants of a study, contextual inquiries also aim to unravel the reasons behind the participants’ actions so as to gain a robust understanding of work practices and behaviours \cite{intro_3}. Since it focuses on in-situ observation of interactions in their natural environment, it seemed well-suited to bring a social viewpoint to bear on social systems \cite{intro_1}, providing an opportunity to ensure the system resonates with the circumstances of its comprising population. 

Contextual inquiry practices overcome the limitations of traditional qualitative research methods like traditional surveys and interviews \cite{intro_6}, it allows for ready chat about what they're doing and why they're doing it while they're doing it. As a result, contextual inquiry can give more detailed and relevant information about how people complete procedures than self-reported or lab-based research. Interestingly, it has been hypothesized that research participants reveal more sensitive and personal information including subjects such as drug abuse, sexual assault experiences and mental health symptoms when they respond anonymously through a questionnaire, rather than through confidential face-to-face or telephonic interviews \cite{intro_ref_3,intro_11,intro_ref_2}.

A feasible solution to obtain explanatory and exploratory data from participants at scale would be the migration of such contextual inquiry procedures to a digitally deliverable platform. This would allow researchers to collect response data including objective responses, subjective opinion and longer textual transcriptions without the involvement of any external being; along with the removal of the participants’ personally identifiable information including personal address information, telephone numbers or any other personal characteristics from the response data. Such a shift, described as crucial in \cite{intro_ref_6} would also improve the present scalability of such research since the utilisation of natural language processing techniques for tasks such as transcription, qualitative data analysis etc. eliminate or at the very least, reduce the need for skilled qualitative researchers and annotators to be involved in the collection, pre-processing as well as the initial analysis phases.

In this work, we propose a framework, ECHO, that drives to meet the above-mentioned exigencies in the contextual inquiry process. In order to support all kinds of response data, similar to that of an in-person inquiry, the proposed framework supports free-form natural language text input to a question in the form of subjective and perceptive semi-structured responses using digital conversational assistants. In parallel, the framework runs algorithms for employing coding methods to analyse qualitative data in real time. Our primary focus is on extraction of meaningful quantitative output from response data, while not compromising on privacy preservation and anonymity of the participant. The framework also supports extensive branching logic, and response driven exploratory questions to enable the querying entity dive deeper into correlation with related questions, and do better cause-effect analysis. 

We understand that subjective natural language answers serve as a double-edged sword — the participants, with guaranteed anonymity, are well-suited to provide us with honest informative data without any inhibitions. However, current systems disallow non-cooperation and lack of interest from the participants for crowd-sourced data contribution, i.e., participants do not have the flexibility to stray from the fixed method of answering questions in the traditional research structure and their responses are generally considered to be candid and consistent. However, in reality, many participants often provide careless and haphazard answers which can compromise the quality of data, thus affecting the reliability of the study’s results. The ECHO framework handles this problem by using NLP techniques to validate the participant responses to ensure consistency, authenticity, and reliability of the participant's responses. With the removal of the interviewer from the process by means of this framework, we not only solve the problem of confidentiality but also remove the possibilities of other challenges such as the interviewer bringing their own biases into the session or even biasing the user - both of which would affect the quality of response data from the participant. Additionally, with the flexibility of providing free-form textual responses, we can capture a larger number of anecdotal responses which serve a greater purpose than objective answers. Through this work, we make the following key contributions:

\begin{itemize}

    \item Design a virtual-assistant driven contextual inquiry framework, ECHO, to eliminate human effort in transcription and interpretation of qualitative response data.
    
    \item Create an information flow pipeline for semi-structured natural language text, from extraction to analyses, using automation of coding methodologies used in qualitative research.  
    
    \item Deploy ECHO on a mental health study on young adults, structured to probe deeper insights into issues faced as well as rationale for inhibition against seeking assistance or resolution.
    
\end{itemize}

\section{Related Work}
\label{sec:related-work}

\subsection{Sensitive Qualitative Studies}

Renzetti and Lee \cite{sensitive_relwork_2} defined sensitive study subjects as topics that scare, discredit, or implicate the participants. The topic being studied could itself be viewed as sensitive, or the exploration and research in the particular topic might evoke emotions in the parties involved. Dickson-Swift et al. \cite{sensitive_relwork_3} interpreted sensitive research as one which could pose as a considerable hazard to individuals who have been or are involved in it, considering that all the stakeholders may be impacted. The effects of such sensitive qualitative studies on its participants was further explored by Bourne \& Robson \cite{sensitive_relwork_11} where they asserted the importance of having a bias-free environment for the participants. It was also observed that the participants experienced cathartic feelings while reflecting on their experiences during the interview  \cite{sensitive_relwork_12,sensitive_relwork_3,sensitive_relwork_10}. 

Sanjari et al. \cite{sanjari2014ethical} explored the limitations of qualitative research and observed that differences in researchers’ skill and training also affected how they evaluated and interpreted the data. Furthermore, Bouchard \cite{sensitive_relwork_13} examined online qualitative research studying sensitive topics, especially the impact of participant anonymity on self-disclosure is specifically which also takes into account the potential drawbacks of conducting qualitative research with participants online, including difficulties in developing a rapport with participants and the researcher, participant authenticity, and participant safety. Anonymity and confidentiality of the participants and the ethics around conducting such sensitive qualitative research also remain an ongoing challenge that is being studied extensively \cite{sensitive_relwork_14}. As evident in aforementioned literature, there exists a trade-off between preserving participant privacy and being able to observe, interpret and draw insights from the participants’ gestures and emotions when the contextual inquiries are conducted in person. ECHO proposed to tread on this precipice by utilizing a conversational assistant framework for contextual inquiry, while extracting insights using automation of conventional qualitative coding methodologies.

\subsection{Coding Methodologies for Qualitative Research}

Over time there have been multiple coding methodologies, each with a distinct purpose, applied to qualitative textual data \cite{relwork_coding_12,relwork_coding_14,relwork_coding_8,relwork_coding_17,relwork_coding_18}. While exploring interpersonal interactions in 1991, Baxter \cite{relwork_coding_11} described thematic analysis (often used interchangeably with terms like ‘content analysis’) as the identification of recurrent themes and patterns in the data. Swain \cite{relwork_coding_12} explored thematic analysis to study 25 semi-structured interviews from a real-life, qualitative study about people’s attitudes toward retirement and their expectations of growing older; providing an excellent example of the application of thematic analysis in ethnography. Glaser and Strauss \cite{relwork_coding_14} pioneered the idea that theoretically important categories and hypotheses can emerge from the observations a qualitative researcher collects (inductive) and even provide answers to the researcher-generated hypotheses (deductive), paving the way for hypothesis coding. Miles et al. \cite{relwork_coding_8} included causal mechanisms as a part of their qualitative research goals, focusing on how and why specific events occurred along with their chronology giving rise to the development of causation coding techniques. Berends et al. \cite{relwork_coding_17} applied such causation techniques to study product innovation processes in small firms by exploring managerial causation. Liu \cite{relwork_coding_18} described emotion coding and sentiment analysis as focusing on extracting emotions from qualitative data and classifying them into predefined categories. The basic framework for axial coding was proposed by Strauss and Corbin \cite{relwork_coding_19} who study the use of a “coding paradigm” to include a variety of factors that influence the phenomenon. It has developed over time to enable researchers to construct linkages between data. Other research methods such as magnitude coding, values coding etc. have evolved also with advancements in qualitative research analysis \cite{relwork_coding_1}. Most of the aforementioned coding methodologies require the researcher to manually extract necessary information from the qualitative data collected. Hence, it is trivial to note that even partial automation of the same would lead to a substantial increase in the overall efficacy of analysing qualitative textual data.

\subsection{Qualitative Data Analysis Tools}

The overarching class of implemented algorithms to study unstructured and semi-structured text, audio or other alternative modality data has been collectively referred to in literature as Computer-Assisted Qualitative Data Analysis Software. White and Rege \cite{white2020sentiment} explored the sentiment analysis service offered by the Google Cloud Platform to examine textual user comment data. The Google Natural Language API  provides powerful pre-built models that can be invoked as a service allowing developers to perform sentiment analysis, along with other features like entity analysis, content classification, and syntax analysis. They found the sentiment analysis service to have an accuracy score of 57\%, largely due a high number of false positives. Pallas et al. \cite{pallas2020evaluating} also provide an overview and comparison of the cloud NLP services offered by Google, Amazon, Microsoft and IBM. They studied sentiment analysis, named entity recognition (NER) and text classification to find significant differences between the providers across examined NLP tasks. Amazon performed the best in sentiment analysis, Google in NER and IBM in text classification. Hwang \cite{hwang2008utilizing} analysed Atlas.ti, a CAQDAS which has proven to be quite useful tool in academic research in the social sciences, emphasising the fact that unlike others, Atlas.ti can handle text, audio, video and other digital media formats. MAXQDA and NVivo \cite{saillard2011systematic} are also quite popular tools for qualitative data analysis.

In the recent years, there has been research on frameworks and tools as well as user experience design which can be leveraged by qualitative researchers. From data collection to employ task designs such as Sprout, where Bragg et al. \cite{uist1} provided an open source framework to efficiently crowdsource data by ensuring cogency to all participants, to Tools such as Screen2words - an automatic mobile UI summarisation tool created by Wang et al. \cite{uist2}, Marcelle - a toolkit designed by Françoise et al. for HCI design \cite{uist3}, Idyll Studio - a structured editor developed by Conlen et al. which enables users to perform basic editing and composition, as well as specify relationships between components \cite{uist4}, and automatic continuous text summarisation developed by Dang et al. \cite{uist5} can be exploited to extract eclectic research data. There have been many developments in the use of conversational assistants as well, such as error message handling in conersational flows \cite{cui5}, GDPR compliance in chatbots \cite{cui4}, psychological approaches to user engagement with chatbots \cite{cui2,cui3} and chatbot dialogue design frameworks \cite{cui1}. The exploratory labelling assistant proposed by Felix et al. \cite{uist6} can also help researchers coherently categorise groups of documents into a set of unknown and evolving labels, thus simplifying the coding process. manipulate and frame more creative research questions \cite{iui1}. This work utilizes some of these tools, as well as prior theoretical frameworks to build an end-to-end pipeline for ECHO.

\begin{figure*}[h]  
    \centering
    \minipage{0.9\textwidth}
        \includegraphics[width=\linewidth]{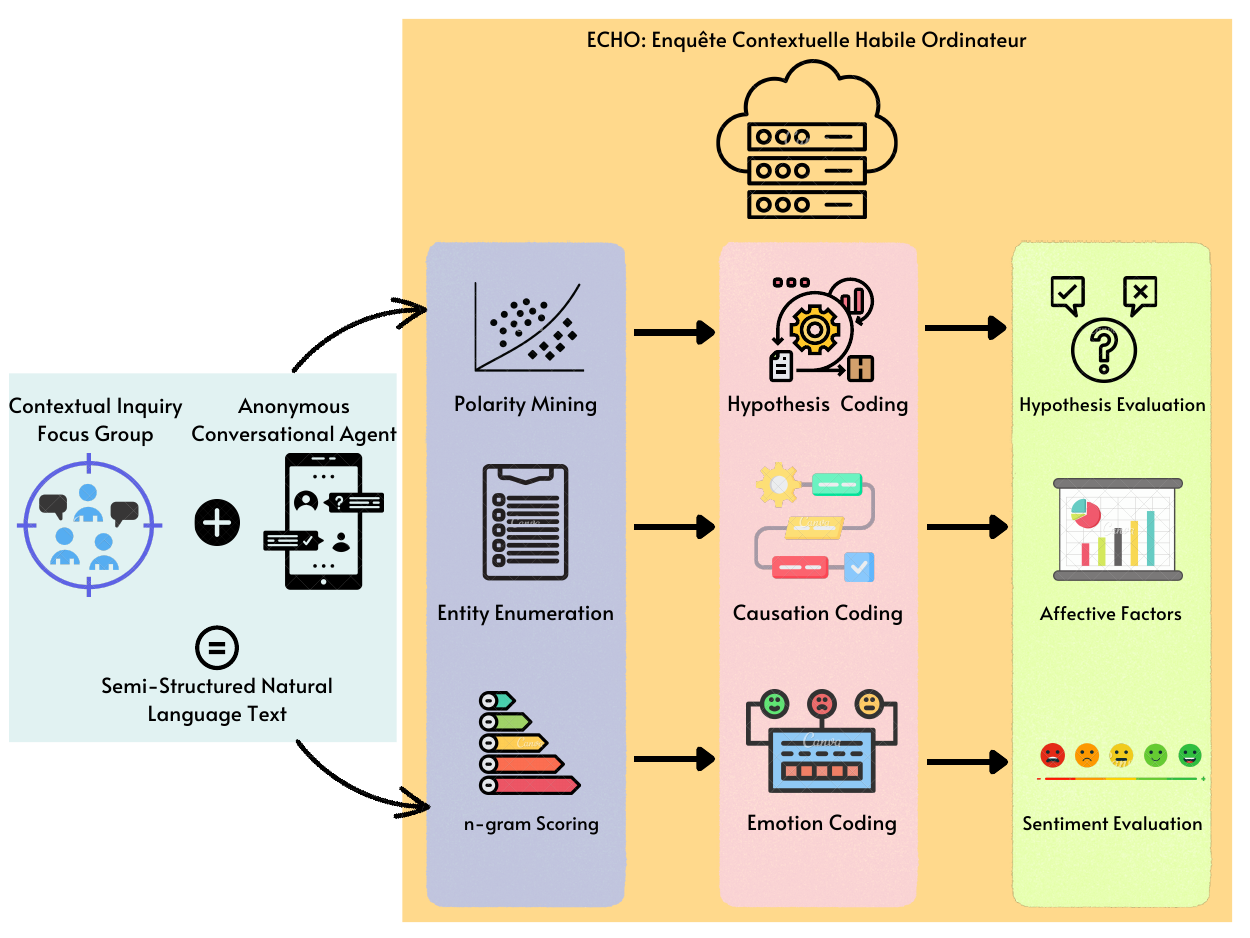}
        \caption{A figurative description for the ECHO framework, with all component layers and information flow, left to right.}
        \label{fig:echo-framework}
    \endminipage
    \hfill
\end{figure*}

\section{System Design}
\label{sec:system-design}
In this section, we will elaborate upon each of the individual components of the ECHO framework, which encumber the end-to-end information processing pipeline of a digital contextual inquiry. When analysing large scale qualitative data, ECHO passes the contextual inquiry (CI) responses through \textit{three} quintessential steps, viz. (a) Quantifying qualitative data, i.e. categorising, labelling, and annotating text to meaningful entities or comparable metric quantities, (b) Automating coding methodologies, thereby helping understand underlying response theme(s) or inquirer hypotheses and their impact on CI outcomes, and (c) Post-procurement analysis using tangible statistical methods that interpret CI results using correlation matrices, frequency CDFs and hypothesis testing. 

To enhance clarity, the qualitative response data fed to the ECHO framework is recorded from anonymous digital contextual inquiry, designed by an inquirer and delivered as conversational dialogue, e.g. a conversational virtual assistant on a smartphone or personal device. Anonymous survey methods appear to promote greater disclosure of sensitive or stigmatizing information compared to non-anonymous methods. Higher disclosure rates have traditionally been interpreted as being more accurate than lower rates. Also, it can be worth discussing that the accuracy or honesty of disclosure for stigmatizing and sensitive personal information might be associated to recruitment and selection of sample population that is affected higher by the experiences under inquiry, rather than simply respondent privacy. With this disclaimer in mind, we proceed to discuss the three operational layers for ECHO. 

\subsection{Quantification of Free-form textual data}
A significant feature of the ECHO framework is the collection of free-form textual data responses which bear a plethora of information, enabling the retrieval of larger narratives that may have been missed otherwise. In order to study such large amounts of qualitative data, it is essential to structure and quantify this data maximally so that we can facilitate a straightforward analysis of the information.

Over 80\% of all qualitative data is unstructured or semi-structured, with textual data being one of the most common types of unstructured data. This makes analysing, understanding, sorting and organising data to draw insights from it, a very difficult and time-consuming task \cite{coding_methods_22}. In order to exploit this unstructured text data to its full potential, the first phase of ECHO employs text-classification and Named Entity Recognition (NER) \cite{goyal2018recent}, thus assigning a list of predetermined categories to open-ended or semi-structured text. Text classifiers have been used to organise, arrange, and categorise natural language data, including text from the web, medical research, and even academic publications \cite{coding_methods_23}. We shall see an empirical use-case for these algorithms in Section \ref{sec:case-study}. From the questions posed in the mental health inquiry of young adults in Table \ref{tab:indices-scoring}, we can identify certain characteristics to be extracted from the textual responses whose quantification would be crucial in aiding further analysis. These components can be generalized and are elaborated below:

\subsubsection{Enumeration of Entities}
ECHO employs entity enumeration or named entity recognition (NER) \cite{coding_methods_26} on the  flexible-length textual responses to such questions. Entity enumeration is a sub-task of information extraction that captures and classifies named entities mentioned in semi-structured responses into pre-defined categories, such as names, places, or expressions common to a central theme.

Similar to text classification, entity extraction from unstructured text can be done manually or automatically via (a) dictionaries and rules, crafted by the inquirer, or (b) supervised learning for entity enumeration \cite{kaiser2005information}. Manual tagging is performed in a similar approach as mentioned before, using a team of annotators and a codebook. Supervised learning approaches are more widely used today, comprising decision trees, hidden Markov models, maximum entropy models, support vector machines, boosted and voted perceptrons, and conditional random fields (CRFs) \cite{coding_methods_27, coding_methods_32}. They generally use semantic, linguistic and knowledge-based extraction methods.

ECHO extracts the unique elements or ‘entities’ present in the responses, along with their relevance to the text block so that we can retain only the suitable and essential components of the text for further analysis. The supported entity extraction retrieves information about the entities present in the text, which typically include proper and common nouns which are returned as indexed offsets into the original text.

\subsubsection{Quantifying frequency of occurrence}
While performing contextual inquiries and other qualitative research, researchers often pose questions regarding the frequency of a particular thought, event or action. For example, “How often do you go to the gym in a week?” or “How many times a year do you visit your grandparents?”. The responses to these questions describe how often this event occurs, in definite (daily, 3 times, 5-6 times etc.) or indefinite terms (frequently, sometimes, once in a while etc.). In the mental health case study conducted via ECHO, we see that Questions 1,6 \& 7 required information about how often the respondents faced trouble sleeping, discussed mental health with their friends, and felt tired. When we receive free form textual responses to such questions, ECHO automates the quantification of the frequency of occurrence of the event being studied, according to a specified scale (weekly, monthly, yearly, etc.). ECHO has certain measures in place which enable it to perform this quantification:

\begin{itemize}
\item \textbf{Format rules for frequency-based questions:} Firstly, the inquirer posing the question via ECHO must abide by a specific structure for the way the question is framed. The questions must provide a certain unit such as ‘days-per-week’ or ‘hours-per-day’, thus making the quantification process easier by providing the ECHO framework with a scale as well as making it cogent to the participants about the type of response that the question seeks. A sample question framed in such a manner is as follows: “On average, how many days a month do you exercise?”. In this case, the activity in focus is exercising and the unit of frequency expected is in ‘days-per-month’. An expected response to this question could be “4 days”. From the participant’s response, we would be able to score it in ‘days-per-week’ (i.e., 4/30.57 or 0.13 days-per-month). If the question posed does not follow this structure, it will not be accepted by the ECHO framework, which will show an error message along with suggestions on how to frame the question in accordance with the necessary format.
    
\item \textbf{Custom NER for frequency-of-occurrence:} To quantize responses, ECHO employs a custom NER recognition process for classification of quantities and relative occurrence of the event in question. We utilised the spaCy v3 \cite{schmitt2019replicable}, a robust open-source NLP framework in Python, to build our custom frequency of occurrence NER model. The pipeline consists of a tagger stage, a parser stage, and an entity recognizer stage.

\smallskip
The training data for expected responses to questions expecting the 'frequency-of-occurrence' of an event or activity needs to be quite comprehensive. The responses may be semi-structured, with the indicated quantity in numbers or words arbitrarily indicated in short or long phrases of natural language text. They may even be dimensionless, e.g. 'twice', ' twice a month', or even, 'fortnightly', which could all be responses to the aforementioned question on exercising. To this end, we prepared a training dataset with relevant frequency-of-occurrence response data, containing authentic medicine and drug reviews from users on WebMD. This user review data contained information in natural language, where adverbs of frequency such as ‘daily’, ‘sometimes’, ‘3 times’ etc. were often used with regards to dosage and its effects. These specific terms and phrases were manually annotated with the entity type ‘frequency-of-occurrence’ for our custom model to identify. This annotated data was converted into .spacy format and fed to the model. As a basis for training our model, ‘en\_core\_web\_lg’ vectors are used. ‘en\_core\_web\_lg’ is a spaCy English multi-task CNN trained on OntoNotes, with GloVe vectors trained on Common Crawl. It assigns word vectors, context-specific token vectors, POS tags, dependency parse and named entities \cite{filgueira2020geoparsing}. spaCy trains the model in an iterative approach, where the gradient of loss is computed by comparing the model predictions with the annotated data. The model weights are then accordingly updated via the backpropagation algorithm, until the model’s predictions gradually resemble the given labels. 

\smallskip
We also devised a custom vocabulary; which is a Python dictionary containing a comprehensive list of adverbs of frequency with a pre-assigned score or a percentage. When the custom NER model classifies a word or phrase as a frequency-of-occurrence entity, it is then compared with the vocabulary dictionary and assigned a score accordingly. For example, an identified frequency-of-occurrence entity “3-5 times” would be regex matched with the key “x-y times'' in the dictionary, where it would be assigned a score of (x+y)/2 or (3+5)/2  i.e., 4. Another entity “daily” would be matched by the dictionary to have a value of 100\%. And, 100\% of a week would be 7 days-per-week which would be the score assigned to the specified response.

\end{itemize}

\subsubsection{Measure of Response Polarity}
There may be questions posed using the ECHO framework where the inquirer is trying to gauge the respondents’ emotion or polarity of opinion towards a particular subject, as attempted by studies in \cite{kim2004determining}. To provide an example for a question of this type posed in the mental health case study in Section \ref{sec:case-study}, we can consider Q6 where we try to assess how comfortable the respondent is with therapy, i.e. studying the polarity in the respondent’s emotions and opinions.

The ECHO framework utilizes pre-trained models offered by the Google Cloud Natural Language API to perform annotation, sentiment analysis, and entity sentiment analysis on the gathered textual responses.

\subsection{Automation of Coding Methodologies}
Coding methodologies are a vital preliminary step in analysing raw data obtained from surveys, audio transcription or secondary sources. Coding techniques help refine and organise qualitative data to identify distinct themes in the data and the relationships between them. 

By quantifying and giving meaning to the pre-processed raw data, they ease the process of data analysis and theme-extraction for later purposes of pattern detection, categorisation, theory building, and other analytic processes \cite{relwork_coding_5}. These coding methodologies in addition to advancements in natural language processing algorithms, drastically enhance the analysis of qualitative survey data to draw meaningful insights and conclusions from them. By employing NLP-based coding approaches such as sentiment polarity, topic extraction, parts-of-speech tagging, relationship extraction, stemming, and more; we can analyse, organise and structure the responses collected in the contextual inquiry \cite{coding_methods_2}. The ECHO framework employs a number of these techniques, viz. sentiment polarity, entity analysis, — to analyse the qualitative data obtained from our contextual inquiry application.

\subsubsection{\textbf{Thematic Coding}}
Thematic analysis is a commonly-used approach in qualitative research studies where the qualitative data is broken down into workable themes to aid in smoother analysis. These are typically esoteric and hard to spot when looking through raw data on its own. Thematic analysis is essential to identify, code, memo and report patterns or themes present within the data, creating a semantic connection between entities that belong to a central major theme within the document. \cite{thematic_1}. 

The training data provided to the classifier model is made up of these pairs of feature sets consisting of vectors for word embeddings for each text sample (TF-IDF vectors or word embeddings such as like GloVe, FastText, and Word2Vec) and tags (such as sports, politics etc.). The model is trained using these features using some of the popular classifiers such as Naive Bayes, SVMs, Boosting Models and Deep neural networks \cite{coding_methods_23}. Inductive thematic coding involves extracting themes directly from the data. This can be done manually or by utilizing machine learning to extract entities from the data and code them into broader themes according to their metadata, syntax or semantics etc.

\subsubsection{\textbf{Hypothesis coding}} 
Hypothesis coding is another important qualitative data analysis method that enables researchers to generate hypotheses and evaluate them using the data. \cite{coding_methods_8}. The results of hypothesis coding allows the inquirer to quantitatively accept or reject a hypothesis using qualitative data and thereafter, quantize response data and draw conclusions from the same. In qualitative data research, hypothesis coding has frequently been used to corroborate or refute any statements, hypotheses, or theories developed \cite{relwork_coding_8}.  

The null hypothesis (H0) generated by the researcher can also be analyzed quantitatively using a number of statistical hypothesis testing techniques such as t-test, F-test etc., enabling researchers to validate the null hypothesis, or reject it for the alternative hypothesis (H1). Such methods allow researchers to identify Type I or Type II errors while performing hypothesis testing. As part of ECHO, in Section \ref{subsec:researcher-hypothesis-evaluation}, we employ a one-sample test on semi-structured responses to evaluate a hypothesis on resistance in young adults to discuss mental health issues with friends or family. Moreover, there exist tests such as ANOVA (Analysis of variance) \cite{case_study_1} and Kruskal-Wallis\cite{mckight2010kruskal}, which enable researchers to make use of the methods’ predetermined hypotheses to further examine their data.

\subsubsection{\textbf{Causation coding}}
Causation coding extracts rationale or causal beliefs from response data, and answers questions such as how and why particular results were derived. This method helps researchers discern motives, belief systems, worldviews etc. in search for causes, conditions, contexts and consequences. It is an important type of narrative analysis that enables researchers to put together the chronological occurrence of events, in addition to their causes. It attempts to map a three-part process as a CODE 1 → CODE 2 → CODE 3 sequence. For example, from qualitative data about cigarette smokers, it could be understood that smoking cigarettes causes cancer and cancer causes lung damage, in that order, or smoking cigarettes → cancer → lung damage \cite{relwork_coding_8}.

\subsubsection{\textbf{Emotion coding}}
Perhaps the most trivial to comprehend, emotion coding in ECHO involves classifying the respondent's emotions and/or sentiments based on semi-structured textual responses. This type of coding takes a dive into the participant’s mentality and perception, enabling the inquirer to assess \cite{coding_methods_8}. 

Sentiment analysis is a series of methods, techniques, and tools about detecting and extracting subjective information, such as opinion and attitudes, from language \cite{coding_methods_16}. Sentiment analysis has traditionally focused on opinion polarity, or whether a person has a positive, neutral, or negative opinion on something. Hence, it serves as a great tool for learning what consumers, in particular, think and feel about a certain topic, idea, or product \cite{coding_methods_9}.

Similar to text classification, sentiment analysis can be performed in 2 ways: manual and automatic. Manual text classification requires a human annotator who analyses the text's content and assigns the appropriate category. Although this procedure can produce good results, it is time and money-consuming.

Two main types of methods exist for automatic sentiment analysis — lexicon models and machine learning models \cite{coding_methods_19}. If we are performing sentiment analysis at a document level, where there is a set of documents D made up of d documents containing opinionated textual data, then we must determine the polarity or orientation expressed in document d about object O having features f1,f2,f3 etc. Using the lexicon approach, we will have a dictionary which is a seed list of words with known orientations before searching internet dictionaries for potential synonyms and antonyms. We would count the number of sentiment words of each category that occur in the textual data, enabling us to classify the sentiment for the object in a particular document. However, creating a dictionary with a relevant set of keywords is a very difficult task since different research questions would require different types of dictionaries. Applying one lexicon or another to look into a certain study subject might result in wildly different findings \cite{coding_methods_20}.

\subsection{Post-Procurement Analytics}

After ECHO has coded the qualitative textual responses in accordance with the coding methodology applicable, the inquirer can utilize the results obtained from this step to furnish tangible outcomes, e.g. hypothesis testing, correlation matrices, etc. in order to study trends, draw conclusions, and make a generalized inference from the data. ECHO implements an array of strategies and techniques for further analysis of processed contextual inquiry data, as applicable to the analysis sought from a particular question-response pair.

\subsubsection{Participant's Response Consistency}
Establishing consistency in the participant's responses throughout the study is crucial. If the participant loses interest or does not answer truthfully, it will affect the dataset any possible conclusions that could be drawn from it. To rule out such inconsistent data, we find correlation among the responses to the different indices. 

Correlation describes the linear association between two quantifiable variables. The degree of correlation is measured by a correlation coefficient called Pearson's correlation coefficient. When one variable increases as the other increases the correlation is positive; when one decreases as the other increases it is negative. Complete absence of correlation is represented by 0. Applying correlation techniques on our data is further explored in section \ref{sec:case-study}.

\subsubsection{Frequency Distributions \& Measures of Central Tendency}
The construction of a frequency distribution, i.e. a structured tabulation or graphical representation of the number of observations in each category on the measurement scale, enables the researcher to glance over all of the data conveniently. It allows researchers to study any trends in the data – if the observations are concentrated at one point or spread out across the entire scale, the range of the data, etc. It provides an overview of the general distribution of the individual observations along the measurement scale \cite{post_proc_1}. Moreover, measures of central tendency such as mean, median, and mode can also be used to represent the entire distribution in a single value \cite{post_proc_2}. An example for the usage of frequency distribution can be seen in the subsequent mental health case study, in section \ref{sec:case-study}.

\subsubsection{Hypothesis Evaluation}
The data from the ECHO framework can be used to perform hypothesis testing and evaluation. Hypothesis testing is used to determine whether a researcher-generated hypothesis is plausible. By performing data analysis, researchers would be able to evaluate and validate their assumptions, enabling them to accept or reject their hypotheses \cite{post_proc_3}. This method can be seen in the subsequent mental health case study in section \ref{sec:case-study}.

\section{Mental Health Case Study}
\label{sec:case-study}

\subsection{Study Structure}

\subsubsection{Participant Demographics and Informed Consent}
We recruited 167 participants to participate in our sample case study exploring the mental well-being of young adults. Our participant demographic primarily comprised university students and young professionals engaging in industry and academia. The study participants were recruited via convenience sampling and participated voluntarily with informed consent. They were made aware of the purpose of the study, data publishing rules and the risks associated with being part of a sensitive data study, prior to their engagement. The study was publicized through curated mailing lists and online support groups for young adults. All participants were between the ages of 17 and 28 years of age (mean age: 20 years, standard deviation: 3.2 years).

\subsubsection{Bifurcated Group Study}
Our goal for the case study was to evaluate the feasibility of ECHO as a contextual inquiry framework in comparison to the conventional contextual inquiry. We wished to investigate the level of detail and the opportunity cost of employing a digital contextual inquiry  in terms of information lost, when compared to a researcher probing for explanatory reasons behind a participant's responses in a traditional scenario. This comparison can be drawn by utilizing identical exploratory question scripts provided via the two media of conducting a contextual inquiry – digital and in-person. Moreover, we wanted to evaluate the virtues of a digitally delivered contextual inquiry, especially when discussing matters which may be extremely sensitive in nature. With the filter of the participants’ personally identifiable information being fully protected (even from the study researchers), they may feel more comfortable revealing details which may be extremely sensitive in nature when compared to an in-person contextual inquiry where participants could be personally identified by the inquirer. In order to review these characteristics, we divided our study participants randomly into two groups where each was given a different medium of contextual inquiry.

\begin{figure*}[h]

    \minipage{0.35\textwidth}
        
        \includegraphics[width=\linewidth]{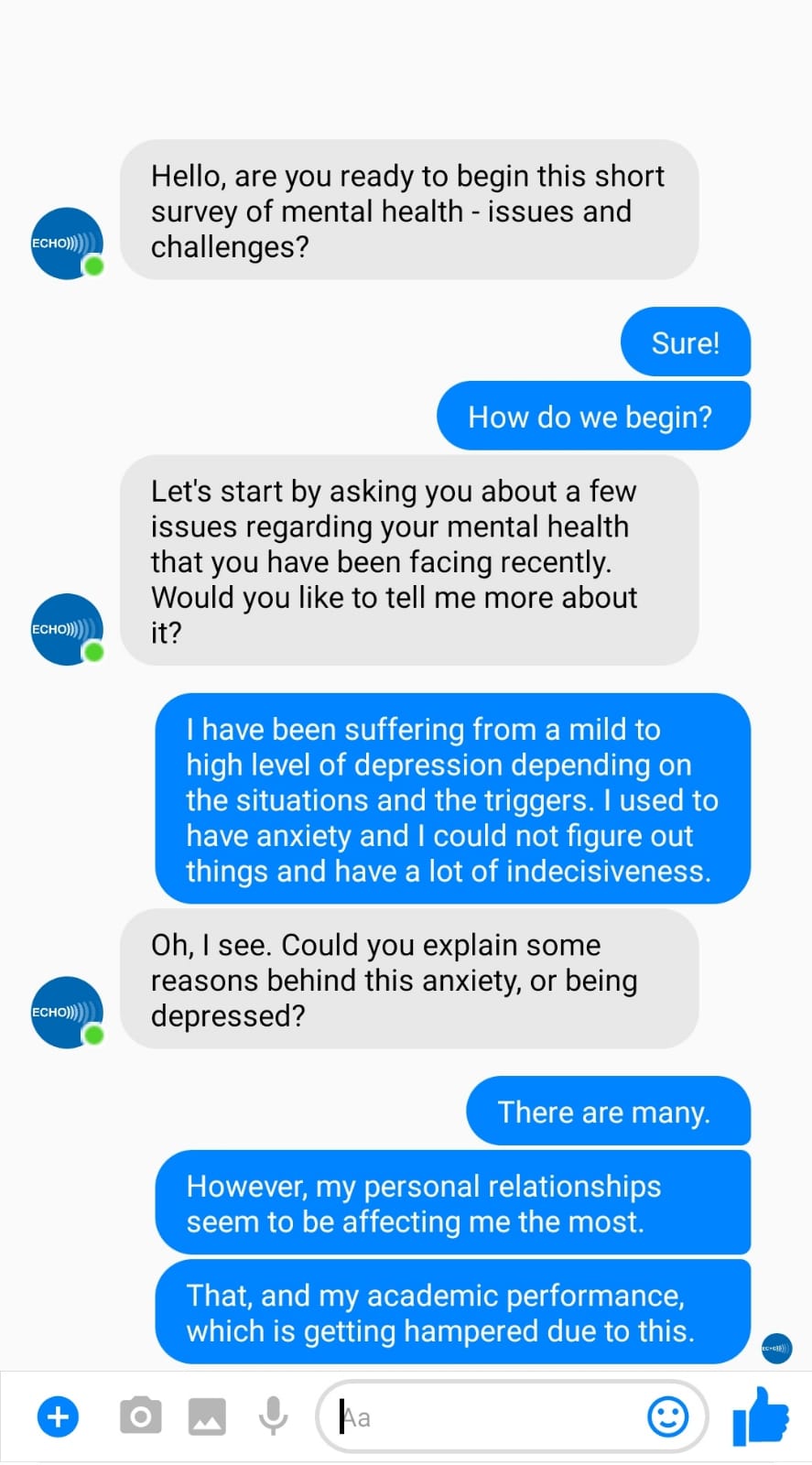}
        
    \endminipage 
    \hfill
    \minipage{0.35\textwidth}
        
        \includegraphics[width=\linewidth]{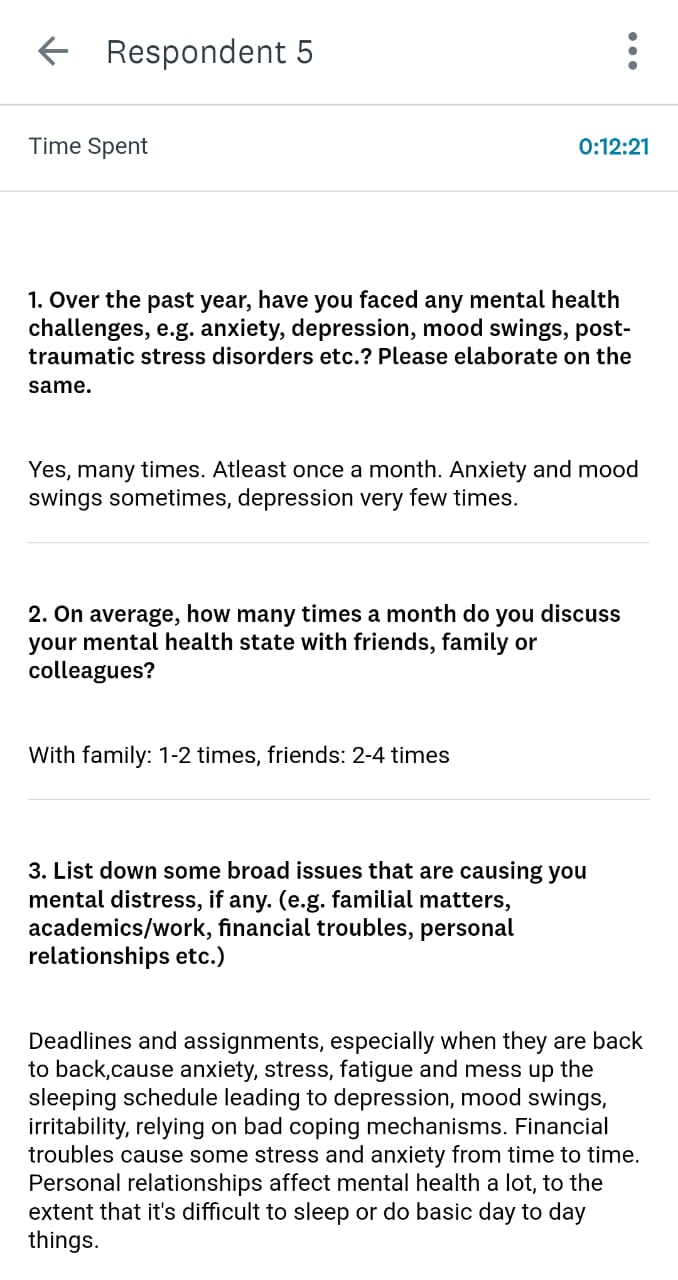}
        
    \endminipage
    \caption{(a) Conversation between ECHO and a respondent, R23, in G2 and (b) Responses extracted from a conversation with a respondent, R5,  in G2}
   \label{figsa}
\end{figure*}

The n=167 sample population (S) of participants were bifurcated into two randomly chosen control groups – Group1 (G1) and Group2 (G2). For G1, we administered an in-person contextual inquiry with the aforementioned questions, where the participant would be inquired by a researcher in an interview-like scenario. The researcher took notes and transcribed the participant’s responses to the probed questions. On the other hand, G2 was given a fully anonymous digitally delivered version of the contextual inquiry, deployed via the ECHO framework, as seen in \ref{figsa} (a). The participants could take part in this study from the comfort of their own environments on a laptop or smartphone with an internet connection, and provide objective or textual responses, as applicable, as seen in \ref{figsa} (b). All the participants were guaranteed total confidentiality and anonymity while they took part in this research study. 

This participatory pilot study utilised objective questions provided by various clinical Mental Health Indices – MHI-5, WHO-9 and PHQ-9 \cite{case_study_2, case_study_3, case_study_4}. These indices collect objective responses from participants about their mental well-being and assign them an index-decided score indicating the state of their perceived mental health. For those participants whose mental health index scores indicated poor mental well-being, we probed in further detail the causes of distress, extent of affecting factors and the rationale of comfort or discomfort in seeking assistance and resolution. 

The qualitative response data gathered was pre-processed by the ECHO framework using multiple coding methodologies, enabling efficient multiple post-procurement analysis techniques on the data to find trends, draw insights and answer researcher-generated hypotheses to further investigate in greater detail the sources of distress and the severity of the influencing elements on the participants.

\subsection{Study Results}
\label{subsec:study-results}

\begin{table*}[t]
\caption{Extracted Entities and Applied Coding Methodologies for the study}
    \label{tab:indices-scoring}
          \centering
	    \begin{tabular}{ccl}
            \toprule 
            
          Question & Entities & Coding Methodology\\
            \midrule
            
             [Q1] Over the past one year, have you &\\how many faced any mental health challenges,&  \\ e.g. anxiety, depression, mood swings, &Factor Enumeration, &Thematic Coding\\post-traumatic stress disorders etc.?  & Affecting Factors \\ Please elaborate on the same. \\
            \hline
          [Q2] On average, how many times a month& \\ do you discuss your mental health state&Frequency of Occurrence& Thematic Coding \\ with friends, family or colleagues?      \\
            \hline
            [Q3] List down some broad issues that &\\are causing you mental distress, if any.& & Emotion Coding\\ (e.g. familial matters, academics/work, & Affecting Factors &Thematic Coding\\financial troubles, personal relationships etc.)\\
            \hline
            [Q4] Have you ever visited or talked to a &\\ psychologist/counsellor regarding/ &Affirmation/Negation&Thematic Coding\\your mental health? \\
            \hline
            [Q5] In your opinion, what are the major &\\factors that deter you from approaching &\\clinical psychologists or counsellors? & Factor Enumeration &Causation Coding  \\ Please elaborate on the same. \\
            \hline
            [Q6] What is your outlook on therapy as a&\\ solution to combat the mental health & \\ issues people face today? Do you think &Opinion Polarity & Emotion Coding\\ solely therapy is adequate to help &\\ people in need? \\
            \hline
            [Q7] What are the prospective feasible factors/&\\ actions that would elevate your mental & Causal Factors & Causation Coding\\health state, given current circumstances?    \\
            
            \bottomrule
           
        \end{tabular}
        

\end{table*}

\subsubsection{Response Consistency}

We analyzed the consistency of the participants’ responses to verify whether respondents were providing sincere responses across all questions. To perform response consistency testing, we chose 3 pairs of questions from the MHI-5 and WHO-5 \cite{case_study_2, case_study_3, case_study_4} mental health indices which must have similar or totally opposite responses. We determined this metric by calculating the correlation coefficients between a participant’s responses to each pair of questions where questions with similar implications would garner similar responses, thus having a strong positive correlation. Similarly, opposite questions would garner opposite responses, having a strong negative correlation value. Based on a participant’s responses to these 3 pairs of questions, we found the correlation between their responses and determined whether they were genuine and consistent throughout the inquiry.  

The following pairs of questions were chosen:
\begin{itemize}
    \item Q1 (WHO-5) \& Q2 (MHI-5), since they must have opposite responses, i.e. a highly negative correlation. 
    \item Q1 (WHO-5) \& Q5 (MHI-5), since they must have highly similar responses, i.e. a strong positive correlation.
    \item Q2 (WHO-5) \& Q3 (MHI-5), since they must have highly similar responses, i.e. a strong positive correlation. 
\end{itemize}

We can take a look at the correlation coefficients between the responses for the chosen questions from a random anonymous participant in Group 1: -0.847 between Q1 of the WHO-5 index \& Q2 of the MHI-5 index; 0.822 between Q1 of the WHO-5 index \& Q5 of the MHI-5 index; and 0.771 between Q2 of the WHO-5 index \& Q3 of the MHI-5 index. These values align with our expected correlation coefficients for these questions, implying that the participant was highly consistent. Similarly, it was found that around 91\% of the respondents were consistent, providing genuine and congruous responses throughout the inquiry process. Moreover, it was observed that 90\% of participants in G1 and 91\% of the participants in G2 were consistent, demonstrating that the consistency of responses was not highly affected by the mode of delivery. Fig {\ref{figsa} (a)} shows a box plot with the variance of the means of the three mental health indices.

\subsubsection{Exploration of Affecting Factors}
Next, we set out to determine the major affecting factors that deterred our respondents from seeking help for their mental health troubles. This was achieved by performing causation and thematic coding. Firstly, we were able to identify broad issues that caused the participants mental distress from the responses to Q3. The transcribed qualitative response data provided by G1 respondents were manually annotated to extract larger themes and narratives. Similarly, we took advantage of ECHO’s entity extraction features for the responses from G2 to identify certain repeating overarching themes that emerged from the qualitative responses gathered. Later, via causation coding, we were able to link the various mental health challenges that the participants faced (Q1) directly to their causes or themes, enabling us to generalise and pinpoint the exact factors that adversely affected the participant’s mental health in a certain manner. It was observed that academics or work related pressure seemed to be the major cause of anxiety for over 77\% of our respondents. Personal relationships and familial matters were found to be primary causes for depression as well. 

Similarly, we also investigated the major factors that deterred students from seeking help for their mental health troubles (Q5). From the responses, we found that over 22\% of the young adult participants stated ‘judgement’ and ‘money’  as their primary affecting factors. Around 9\% of the respondents also mentioned that they were uncomfortable opening about their issues to ‘strangers’. From the word cloud in Fig {\ref{figsa} (b)} we can see participants state many other factors such as ‘time’, privacy, ‘stress’ and ‘anxiety’. Furthermore, while comparing the responses to Q1,Q3 and Q5 between the G1 and G2 participants, we noticed that the G2 participants provided more in-depth and intimate factors in their responses as compared to the G1 respondents, who provided more vague and generic answers. G2 respondents also opened up more about their personal and familial relationship matters, providing finer details and context, whereas majority of the G1 respondents were found to have a reserved and taciturn composure while discussing such topics.

\subsubsection{Participant Opinion Polarity}
We analyzed the sentiment polarity of the responses to Q6, to explore the participants' opinions towards therapy. In addition to transcription, the G1 inquirers also had to assign whether the participant’s feelings were Positive. Negative, or Neutral; assigning them a score based on their response. On the other hand, sentiment analysis was performed via ECHO on the G2 responses to analyse participant opinion polarity. It was found that 32\% of the respondents had an average sentiment score of -0.3 with a normalised magnitude score of 0.09, indicating a mostly neutral (but leaning towards negative) feeling towards therapy being solely adequate for mental well-being. Most of these respondents discussed various other factors such as exercise, self care, quality time with friends and family etc. which provided a more holistic solution to mental health troubles faced by the youth today. Around 17\% of the respondents also had a positive sentiment score of +0.4 where they spoke about the many positives of therapy. 

In G1, the sentiment polarity measured by the researcher is subjective, based on the inquirer accurately picking up on the nuances in the participant’s responses. With the inquirers having different levels of skill and training, it would be impossible to avoid researcher bias and have consistent or generalised metrics to classify the polarity of the responses. But such challenges were not faced in G2, since ECHO performed algorithmic sentiment analysis to study the provided textual data, assigning each response a sentiment score which tells us whether the respondent had a positive, negative or neutral opinion. We also observed that many G2 participants provided examples from their personal experiences with therapy to back their responses, while G1 participants very rarely revealed such information.

\begin{figure*}[h] 
\begin{center}
    \begin{minipage}[t]{0.35\textwidth}
        \includegraphics[width=\linewidth]{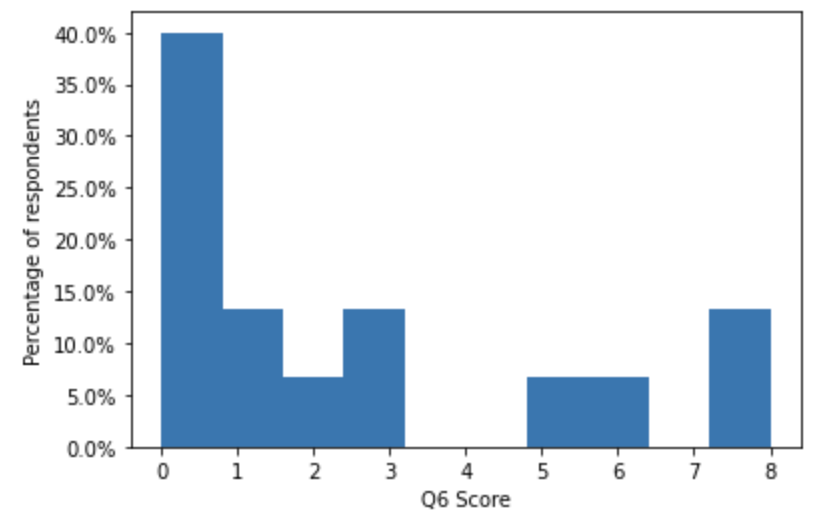}    
    \end{minipage}
    \begin{minipage}[t]{0.35\textwidth}
        \includegraphics[width=\linewidth]{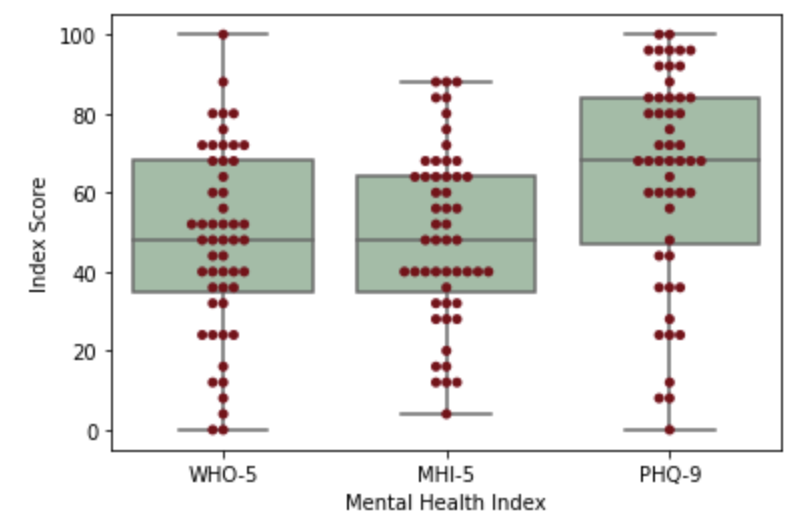}
    \end{minipage}
    \caption{(a) Frequency distribution for the responses to Q6, and (b) Box plot of WHO-5, MHI-5, and PHQ-9}
    \label{figsb}
    \end{center}
\end{figure*}

\subsubsection{Researcher Hypothesis Evaluation}
\label{subsec:researcher-hypothesis-evaluation}
Based on our literature review and preliminary analysis, we developed a hypothesis H0 which states that in spite of having mental health challenges, young adults today do not take the step to seek help from clinical mental health professionals and psychologists (Q4), or discuss it openly with their friends and family (Q2). From our exploratory study, we found that around 71\% of the respondents facing mental health troubles had never visited a psychologist or counsellor for their mental health. Moreover, 35\% of our respondents had also never had an open conversation about their mental health with their friends or loved ones. From our findings, it is evident that most of the young adults from our sample population had never sought external help for their mental health, thus validating our hypothesis H0.    

We also observed that the number of participants who admitted to seeking help in G2 was 9.7\% higher than in G1. A probable cause for this phenomenon could be that respondents felt more comfortable about revealing the fact that they went to therapy in an anonymous mode, rather than in front of someone else (i.e., the inquirer).

\section{Discussions}
\label{sec: discussion}
\subsection{Lessons from the Exploratory Study}
From our study results, it is evident that the level of detail parity between the G1 and G2 responses were notable – respondents from G2 provided more detailed and collected responses to the questions, as compared to G1. This suggests that the anonymous digitally delivered method does perform better than the conventional contextual inquiry, in this context. We have also seen that people felt more comfortable sharing information of extremely sensitive or intimate nature more confidently in a digital mode rather than in-person. Thus, virtues of such a digitally delivered anonymous mode of contextual inquiry become apparent. In addition to the aforementioned benefits, such a framework also makes participation in such studies easily accessible to members from varied geographic, economic and educational backgrounds, from the convenience of their own environment.

\subsection{Ethical Considerations}
While conducting this research, multiple ethical implications of conducting digital inquiry were found. Epistemic concerns, especially inconclusive evidence and misguided evidence were identified as possible ethical considerations in deploying algorithms to assess sensitive data. Algorithmic decision making often relies on correlations within a dataset. Causality is frequently looked over, perhaps because searching for causal links is a highly difficult process, since studies on large datasets often yield results which are not usually reproducible \cite{lazer2014parable}. Any actions taken on the basis of correlations can be problematic when causality has not been established. To combat this issue of inconclusive evidence, we ensured that no actions were taken on the basis of our correlations. Correlation algorithms were used specifically to establish a correlation between the mental health indices. Any actions that would involve prior establishment of causality were avoided.
Conclusions drawn in a qualitative study can only be as reliable and neutral as the data its based on. The belief that algorithms lack bias has been debunked in several studies\cite{newell2015strategic}. The values of the developer are inevitably behind an algorithm's design, which can lead to questionable neutrality. The outputs of our algorithm also require interpretation, which can again be highly subjective. To ensure that the results accurately reflect the responses of our respondents rather than the biases of the interpreter, we ensured that there were more than one person working on the design of the algorithm as well as the interpretation of the results. Each had the power to veto other's responses until a common consensus could be reached regarding the    interpretation, thus reducing chances of any bias and misguided evidence.

\subsection{Limitations and Challenges}
Anonymization of respondents is considered a default position on ethical grounds in several qualitative studies, including ours. Surveys dealing with personally identifiable data need to ensure that respondents are not identifiable and should not suffer harm as a consequence of the research. While digitizing qualitative research certainly aids anonymity, there were still certain limitations. Anonymity was difficult to achieve since our study focused on young adults whom the interviewers were familiar with. While measures were taken to ensure that respondents known to the interviewers refrained from participating, it is not a method guaranteed to yield completely anonymity. 

Additionally, digitizing contextual inquiry poses its own set of challenges. Online data collection reduces the burden of time and cost for respondents to participate in the research as well as move beyond the limits posed by geographical barriers. However, digitizing contextual inquiry risks alienating a demographic which is not comfortable using technology and internet, or who might not have the resources to do so. Additionally, not all technology is inclusive of every disability, which makes accommodating participants of every background practically impossible. In such situations, traditional methods of conducting contextual inquiry seem to fair better. 

In-person inquiry also fairs better when reading the respondent's expression and body language is required. Analysing non-verbal cues gives further insight to a respondent's behaviour during in-person inquiry which is not possible in digital inquiry. The latter is also incapacitated when the participants become non-compliant or lose interest because of the lack of human reactions and engagement which are present in in-person inquiry.

\section{Conclusion and Future Work}
\label{sec:conclusion-and-future-work}
In this paper, we have presented a framework for digital contextual inquiry. We have proposed Enquête Contextuelle Habile Ordinateur (ECHO), a framework designed for (a) privacy preservation and anonymity of the participant by eliminating any mention of personally identifiable information, (b) flexibility of input responses allowed for contextual inquiry and (c) removal of manual methods to transcribe and interpret data during the contextual inquiry process. 
For future work, we aim to implement and test multi-modal contextual inquiry, and implement our model in certain location-based contexts, e.g.deliver to participants on sensitive workplace issues - gender-sensitive facilities, bullying, flouted rules or co-worker ethics, racial discrimination etc. Ensuring anonymity would potentially make the framework adapt to a relatively charged atmosphere.

\bibliographystyle{ACM-Reference-Format}
\bibliography{sample-base}

\end{document}